\documentstyle[11pt,aaspp4]{article}
\begin{document}

\title{DIRECT Distances to Nearby Galaxies Using Detached Eclipsing 
Binaries and Cepheids. II. Variables in the Field M31A\footnote{Based
on the observations collected at the Michigan-Dartmouth-MIT (MDM)
1.3-meter telescope and at the F.~L.~Whipple Observatory (FLWO)
1.2-meter telescope}}

\author{K. Z. Stanek\altaffilmark{2}} 
\affil{\tt e-mail: kstanek@cfa.harvard.edu} 
\affil{Harvard-Smithsonian Center for Astrophysics, 60 Garden St., MS20, 
Cambridge, MA~02138} 
\altaffiltext{2}{On leave from N.~Copernicus Astronomical Center, 
Bartycka 18, Warszawa 00--716, Poland} 
\author{J. Kaluzny}
\affil{Warsaw University Observatory, Al. Ujazdowskie 4,
00--478 Warszawa, Poland} 
\affil{\tt e-mail: jka@sirius.astrouw.edu.pl} 
\author{M. Krockenberger, D. D. Sasselov} 
\affil{\tt e-mail: krocken@cfa.harvard.edu, sasselov@cfa.harvard.edu} 
\affil{Harvard-Smithsonian Center for Astrophysics, 60 Garden St., MS16,
Cambridge, MA~02138} 
\author{J. L. Tonry} 
\affil{University of Hawaii, Institute for Astronomy, 
2680 Woodlawn Dr., Honolulu, HI~96822}
\affil{\tt e-mail: jt@avidya.ifa.hawaii.edu} 
\author{M. Mateo}
\affil{Department of Astronomy, University of Michigan, 
821 Dennison Bldg., Ann Arbor, MI~48109--1090} 
\affil{\tt e-mail: mateo@astro.lsa.umich.edu}

\begin{abstract}

We undertook a long term project, DIRECT, to obtain the direct
distances to two important galaxies in the cosmological distance
ladder -- M31 and M33, using detached eclipsing binaries (DEBs) and
Cepheids. While rare and difficult to detect, detached eclipsing
binaries provide us with the potential to determine these distances
with an accuracy better than 5\%. The massive photometry obtained in
order to detect DEBs provides us with the light curves for the Cepheid
variables. These are essential to the parallel project to derive
direct Baade-Wesselink distances to Cepheids in M31 and M33. For both
Cepheids and eclipsing binaries the distance estimates will be free of
any intermediate steps.
 
As a first step of the DIRECT project, between September 1996 and
January 1997 we have obtained 36 full nights on the
Michigan-Dartmouth-MIT (MDM) 1.3-meter telescope and 45 full/partial
nights on the F. L. Whipple Observatory (FLWO) 1.2-meter telescope to
search for detached eclipsing binaries and new Cepheids in the M31 and
the M33 galaxies.  In this paper, second in the series, we present the
catalog of variable stars, most of them newly detected, found in the
field M31A ($\alpha_{2000.0},\delta_{2000}=11.34\deg,41.73\deg$).  We
have found 75 variable stars: 15 eclipsing binaries, 43 Cepheids and
17 other periodic, possible long period or non-periodic variables. The
catalog of variables, as well as their photometry and finding charts,
are available using the {\tt anonymous ftp} service and the {\tt WWW}.
The CCD frames are available on request.

\end{abstract}

\keywords{distance scale---galaxies:individual(M31,M33)---eclipsing
binaries---Cepheids}

\section{Introduction}

The two nearby galaxies -- M31 and M33, are stepping stones to most of
our current effort to understand the evolving universe at large
scales.  First, they are essential to the calibration of the
extragalactic distance scale (Jacoby et al.~1992; Tonry et
al.~1997). Second, they constrain population synthesis models for
early galaxy formation and evolution, and provide the stellar
luminosity calibration. There is one simple requirement for all this
-- accurate distances.
 
Detached eclipsing binaries (DEBs) have the potential to establish
distances to M31 and M33 with an unprecedented accuracy of better than
5\% and possibly to better than 1\%. These distances are now known to
no better than 10-15\%, as there are discrepancies of $0.2-0.3\;{\rm
mag}$ between RR Lyrae and Cepheids distance indicators (e.g.~Huterer,
Sasselov \& Schechter 1995).  Detached eclipsing binaries (for reviews
see Andersen 1991, Paczy\'nski 1997) offer a single step distance
determination to nearby galaxies and may therefore provide an accurate
zero point calibration -- a major step towards very accurate
determination of the Hubble constant, presently an important but
daunting problem for astrophysicists (see the papers from the recent
``Debate on the Scale of the Universe'': Tammann 1996, van den Bergh
1996).
 
The detached eclipsing binaries have yet to be used (Huterer et
al.~1995; Hilditch 1996) as distance indicators to M31 and
M33. According to Hilditch (1996), there are about 60 eclipsing
binaries of all kinds known in M31 (Gaposchkin 1962; Baade \& Swope
1963; Baade \& Swope 1965) and only {\em one} in M33 (Hubble 1929)!
Only now does the availability of large format CCD detectors and
inexpensive CPUs make it possible to organize a massive search for
periodic variables, which will produce a handful of good DEB
candidates. These can then be spectroscopically followed-up with the
powerful new 6.5-10 meter telescopes.

The study of Cepheids in M31 and M33 has a venerable history (Hubble
1926, 1929; Gaposchkin 1962; Baade \& Swope 1963; Baade \& Swope
1965). In the 80's Freedman \& Madore (1990) and Freedman, Wilson \&
Madore (1991) studied small samples of the earlier discovered
Cepheids, to build PL relations in M31 and M33, respectively.
However, both the sparse photometry and the small samples do not
provide a good basis for obtaining direct Baade-Wesselink distances
(e.g. Krockenberger, Sasselov \& Noyes 1997) to Cepheids -- the need
for new digital photometry has been long overdue. Recently, Magnier et
al.~(1997) surveyed large portions of M31, which have previously been
ignored, and found some 130 new Cepheid variable candidates.  Their
light curves are however rather sparsely sampled and in $V$ band only.

In Kaluzny et al.~(1998) (hereafter: Paper I), the first paper of the
series, we presented the catalog of variable stars found in one of the
fields in M31, called M31B. Here we present the catalog of variables
from the neighboring field M31A. In Sec.2 we discuss the selection of
the fields in M31 and the observations. In Sec.3 we describe the data
reduction and calibration. In Sec.4 we discuss briefly the automatic
selection we used for finding the variable stars. In Sec.5 we discuss
the classification of the variables.  In Sec.6 we present the catalog
of variable stars.

\section{Fields selection and observations}

M31 was primarily observed with the McGraw-Hill 1.3-meter telescope at
the MDM Observatory. We used the front-illuminated, Loral $2048^2$ CCD
Wilbur (Metzger, Tonry \& Luppino 1993), which at the $f/7.5$ station
of the 1.3-meter has a pixel scale of $0.32\;arcsec/pixel$ and field
of view of roughly $11\;arcmin$. We used Kitt Peak Johnson-Cousins
$BVI$ filters.  Some data for M31 were also obtained with the
1.2-meter telescope at the FLWO, where we used ``AndyCam'' with
thinned, back-side illuminated, AR coated Loral $2048^2$ CCD.  The
pixel scale happens to be essentially the same as at the MDM 1.3-meter
telescope. We used standard Johnson-Cousins $BVI$ filters.

Fields in M31 were selected using the MIT photometric survey of M31 by
Magnier et al.~(1992) and Haiman et al.~(1994) (see Paper I, Fig.1).
We selected six $11'\times11'$ fields, M31A--F, four of them (A--D)
concentrated on the rich spiral arm in the north-eastern part of M31,
one (E) coinciding with the region of M31 searched for microlensing by
Crotts \& Tomaney (1996), and one (F) containing the giant star
formation region known as NGC206 (observed by Baade \& Swope
1963). Fields A--C were observed during September and October 1996
5--8 times per night in the $V$ band, resulting in total of 110-160
$V$ exposures per field. Fields D--F were observed once a night in the
$V$-band. Some exposures in $B$ and $I$ bands were also taken. M31 was
also occasionally observed at the FLWO 1.2-meter telescope, whose main
target was M33.

In this paper we present the results for the M31A field.  We obtained
for this field useful data during 29 nights at the MDM, collecting a
total of 109 exposures of $900\;sec$ in $V$, 27 exposures of
$600\;sec$ in $I$ and 2 exposures of $1200\;sec$ in $B$. We also obtained
for this field useful data during 15 nights at the FLWO, collecting a
total of 8 exposures of $900\;sec$ in $V$ and 18 exposures of
$600\;sec$ in $I$. The complete list of exposures for this field and
related data files are available through {\tt anonymous ftp} on {\tt
cfa-ftp.harvard.edu}, in {\tt pub/kstanek/DIRECT} directory. Please
retrieve the {\tt README} file for instructions.  Additional
information on the DIRECT project is available through the {\tt WWW}
at {\tt http://cfa-www.harvard.edu/\~\/kstanek/DIRECT/}.

\section{Data reduction, calibration and astrometry}

The details of the reduction procedure are given in Paper I.
Preliminary processing of the CCD frames was done with the standard
routines in the IRAF-CCDPROC package.\footnote{IRAF is distributed by
the National Optical Astronomy Observatories, which are operated by
the Associations of Universities for Research in Astronomy, Inc.,
under cooperative agreement with the NSF} Stellar profile photometry
was extracted using the {\it Daophot/Allstar} package (Stetson 1987,
1991).  We selected a ``template'' frame for each filter using a
single frame of particularly good quality.  These template images were
reduced in a standard way (Paper I).  Other images were reduced using
{\it Allstar} in the fixed-position-mode using as an input the
transformed object list from the template frames.  For each frame the
list of instrumental photometry derived for a given frame was
transformed to the common instrumental system of the appropriate
``template'' image.  Photometry obtained for the $V$ \& $I$ filters was
combined into separate data bases. Unlike for the M31B field, M31A
images obtained at the FLWO were reduced using MDM ``templates''.

To obtain the photometric calibration we observed 4 Landolt (1992)
fields containing a total of 18 standards stars. These fields were
observed through $BVI$ filters at air-masses ranging from 1.2 to
1.70. The transformation from the instrumental to the standard system
was derived and described in Paper I.  The derived transformation
satisfactorily reproduces the $V$ magnitudes and $V-I$ colors. The
$B-V$ transformation reproduces the standard system poorly, and we
decided to drop the $B$ data from our analysis, especially since we
took only 2 $B$ frames.

To check the internal consistency of our photometry we compared the
photometry for 20 $V<20$ and 47 $I<20$ common stars in the overlap
region between the fields M31A and M31B (Paper I, Fig.1).  There was
an offset of $0.022\;{\rm mag}$ in $V$ and $0.018\;{\rm mag}$ in $I$,
i.e. well within our estimate of the $0.05\;mag$ systematic error
discussed in Paper I.  We also derived equatorial coordinates for
all objects included in the data bases for the $V$ filter. The
transformation from rectangular coordinates to equatorial coordinates
was derived using $\sim200$ stars identified in the list published by
Magnier et al.~(1992).

\section{Selection of variables}

\begin{figure}[t]
\plotfiddle{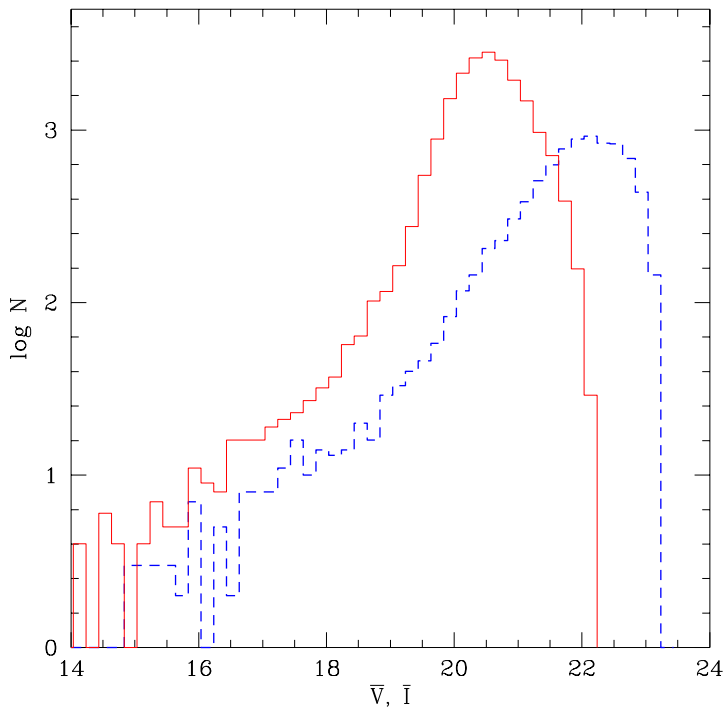}{8cm}{0}{50}{50}{-160}{-85}
\caption{Distributions in $V$ (dashed line) and $I$ (continuous line)
of stars in the field M31A.}
\label{fig:dist}
\end{figure}

The procedure for selecting the variables was described in detail in
Paper I, so here we only give a short description, noting changes when
necessary.  The reduction procedure described in previous section
produces databases of calibrated $V$ and $I$ magnitudes and their
standard errors. The $V$ database for M31A field contains 10084 stars
with up to 117 measurements, and the $I$ database contains 21341 stars
with up to 45 measurements. Fig.\ref{fig:dist} shows the distributions
of stars as a function of mean $\bar{V}$ or $\bar{I}$ magnitude.  As
can be seen from the shape of the histograms, our completeness starts
to drop rapidly at about $\bar{V}\sim22$ and $\bar{I}\sim20.5$. The
primary reason for this difference in the depth of the photometry
between $V$ and $I$ is the level of the combined sky and background
light, which is about three times higher in the $I$ filter than in the
$V$ filter.

The measurements flagged as ``bad'' and measurements with errors
exceeding the average error by more than $4\sigma$ are removed
(Paper~I).  Usually 0--10 points are removed, leaving the majority of
stars with roughly $N_{good}\sim105-117$ $V$\/ measurements.  For
further analysis we use only those stars which have at least
$N_{good}>N_{max}/2\;(=58)$ measurements. There are 8521 such stars in
the $V$ database of the M31A field.

\begin{figure}[t]
\plotfiddle{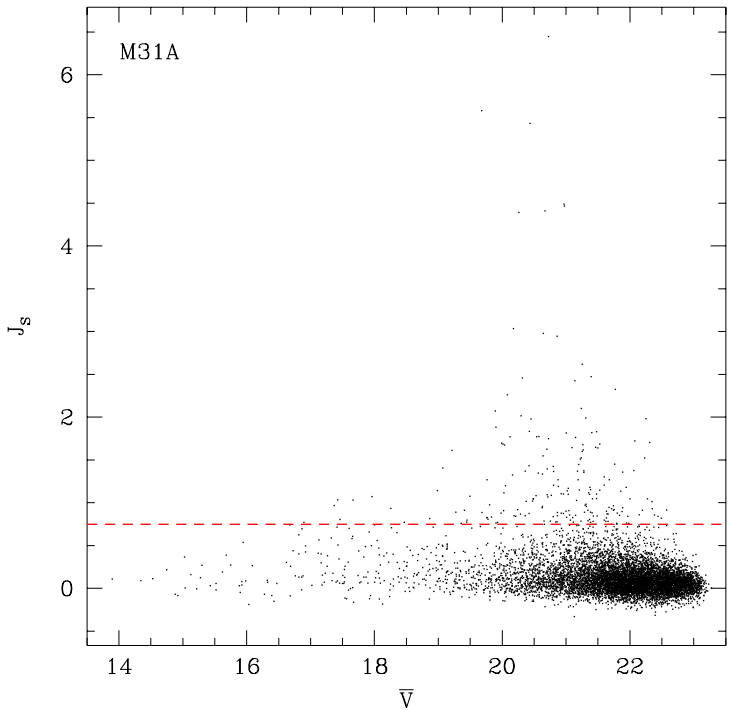}{8cm}{0}{50}{50}{-160}{-85}
\caption{Variability index $J_S$ vs. mean $\bar{V}$ magnitude for 8521
stars in the field M31A with $N_{good}>58$.  Dashed line at $J_S=0.75$
defines the cutoff applied for variability.}
\label{fig:stetj}
\end{figure}

Our next goal is to select objectively a sample of variable stars from
the total sample defined above.  There are many ways to proceed, and
we largely follow the approach of Stetson (1996).  The procedure is
described in more detail in Paper~I. In short, for each star we
compute the Stetson's variability index $J_S$ (Paper I, Eq.7), and
stars with values exceeding some minimum value $J_{S,min}$ are
considered candidate variables.  The definition of Stetson's
variability index includes the standard errors of individual
observations.  If, for some reason, these errors were over- or
underestimated, we would either miss real variables, or select
spurious variables as real ones. Using the procedure described in
Paper I, we scale the {\em Daophot} errors to better represent the
``true'' photometric errors.  We then select the candidate variable
stars by computing the value of $J_S$ for the stars in our $V$
database.  We used a cutoff of $J_{S,min}=0.75$ and additional cuts
described in Paper I to select 183 candidate variable stars (about 2\%
of the total number of 8521).  In Fig.\ref{fig:stetj} we plot the
variability index $J_S$ vs. apparent visual magnitude $\bar{V}$ for
8521 stars with $N_{good}>58$.

\section{Period determination, classification of variables}

We based our candidate variables selection on the $V$ band data
collected at the MDM and the FLWO telescopes.  We also have the $I$
band data for the field, up to 27 MDM epochs and up to 18 FLWO
epochs. As mentioned above, the $I$ photometry is not as deep as the
$V$ photometry, so some of the candidate variable stars do not have an
$I$ counterpart. We will therefore not use the $I$ data for the period
determination and broad classification of the variables. We will
however use the $I$ data for the ``final'' classification of some
variables.

Next we searched for the periodicities for all 183 candidate
variables, using a variant of the Lafler-Kinman (1965) technique
proposed by Stetson (1996). Starting with the minimum period of
$0.25\;days$, successive trial periods are chosen so
\begin{equation}
P_{j+1}^{-1}=P_{j}^{-1}-\frac{0.01}{\Delta t},
\end{equation}
where $\Delta t=t_{N}-t_{1}$ is the time span of the series.  The
maximum period considered is $\Delta t=55.8\;days$.  For each
candidate variable 10 best trial periods are selected (Paper I) and
then used in our classification scheme.

The variables we are most interested in are Cepheids and eclipsing
binaries (EBs). We therefore searched our sample of variable stars for
these two classes of variables. As mentioned before, for the broad
classification of variables we restricted ourselves to the $V$ band
data.  We will, however, present and use the $I$ band data, when
available, when discussing some of the individual variable stars.

In the search for Cepheids we followed the approach by Stetson (1996)
of fitting template light curves to the data. We used the
parameterization of Cepheid light curves in the $V$ band as given by
Stetson (1996). Unlike for the M31B field, we classified the star as a
Cepheid if the reduced $\chi^2/N_{DOF}$ of the fit was factor of 3
smaller than the reduced $\chi^2/N_{DOF}$ of a straight line fit, not
a factor of 2 smaller. We also allowed for periods longer than
$3\;days$, not longer than $4\;days$ as in Paper I, which resulted in
finding two possible Cepheids with periods $3<P<4\;days$.  There was a
total of 51 variables passing all of the criteria.  Their parameters
and light curves are presented in the Sections~6.2,~6.3.

For eclipsing binaries (EBs) we used similar search strategy to that
described in detail in Paper I, but we simplified the condition for
the reduced $\chi^2/N_{DOF}$ of the fit to be at least 1.75 smaller
than the reduced $\chi^2/N_{DOF}$ of a straight line fit. Within our
assumption the light curve of an EB is determined by nine parameters:
the period, the zero point of the phase, the eccentricity, the
longitude of periastron, the radii of the two stars relative to the
binary separation, the inclination angle, the fraction of light coming
from the bigger star and the uneclipsed magnitude.  A total of 15
variables passing all of the criteria and their parameters and light
curves are presented in the Section~6.1.

After we selected 15 eclipsing binaries and 51 possible Cepheids, we
were left with 118 ``other'' variable stars.  After raising the
threshold of the variability index to $J_{S,min}=1.2$ (Paper I) we are
left with 29 variables, which we preliminary classify as
``miscellaneous''.  We then go back to the CCD frames and try to see
by eye if the inferred variability is indeed there, especially in
cases when the light curve is very noisy/chaotic.  We decided to
remove 2 dubious Cepheids and 18 dubious miscellaneous variables from
the sample, which leaves 11 variables which we classify as
miscellaneous. Their parameters and light curves are presented in the
Section~6.4.

\section{Catalog of variables}

In this section we present light curves and some discussion of the 75
variable stars discovered in our survey.  Complete $V$ and (when
available) $I$ photometry and $128\times128\;pixel$ ($\sim
40''\times40''$) $V$ finding charts for all variables are available
through the {\tt anonymous ftp} on {\tt cfa-ftp.harvard.edu}, in {\tt
pub/kstanek/DIRECT} directory. Please retrieve the {\tt README} file
for the instructions and the list of files. These data can also be
accessed through the {\tt WWW} at the {\tt
http://cfa-www.harvard.edu/\~\/kstanek/DIRECT/}.  The CCD frames for
this field, and also for the M31B field, are available on request from
K. Z. Stanek ({\tt kstanek@cfa.harvard.edu}).

The variable stars are named according to the following convention:
letter V for ``variable'', the number of the star in the $V$ database,
then the letter ``D'' for our project, DIRECT, followed by the name of
the field, in this case (M)31A, e.g. V3407 D31A.
Tables~\ref{table:ecl}, \ref{table:ceph}, \ref{table:per} and
\ref{table:misc} list the variable stars sorted broadly by four
categories: eclipsing binaries, Cepheids, other periodic variables and
``miscellaneous'' variables, in our case meaning ``variables with no
clear periodicity''. Some of the variables which were found
independently by survey of Magnier et al.~(1997) are denoted in the
``Comments'' by ``Ma97 ID'', where the ``ID'' is the identification
number assigned by Magnier at al.~(1997). We also identify several
variables found by us in Paper I.

Note that this is a first step in a long-term project and we are
planning to collect additional data and information of various kind
for this and other fields.  As a result, the current catalog might
undergo changes, due to addition, deletion or re-classification of
some variables.

\subsection{Eclipsing binaries}

In Table~\ref{table:ecl} we present the parameters of the 15 eclipsing
binaries in the M31A field.  The lightcurves of these variables are
shown in Figs.\ref{fig:ecl1}--\ref{fig:ecl3}, along with the simple
eclipsing binary models discussed in the Paper I.  The variables are
sorted in the Table~\ref{table:ecl} by the increasing value of the
period $P$. For each eclipsing binary we present its name, 2000.0
coordinates (in degrees), value of the variability index $J_S$, period
$P$, magnitudes $V_{max}$ and $I_{max}$ of the system outside of the
eclipse, and the radii of the binary components $R_1,\;R_2$ in the
units of the orbital separation.  We also give the inclination angle
of the binary orbit to the line of sight $i$ and the excentricity of
the orbit $e$. The reader should bear in mind that the values of
$V_{max},\;I_{max},\; R_1,\;R_2,\;i$ and $e$ are derived with a
straightforward model of the eclipsing system, so they should be
treated only as reasonable estimates of the ``true'' value.

One of the eclipsing binaries found, V4636 D31A, is a good DEB
candidate, with reasonably deep eclipses and the ellipticity
indicating that the system is young and unevolved. However, much
better light curve is necessary to accurately establish the properties
of the system.

\begin{figure}[p]
\plotfiddle{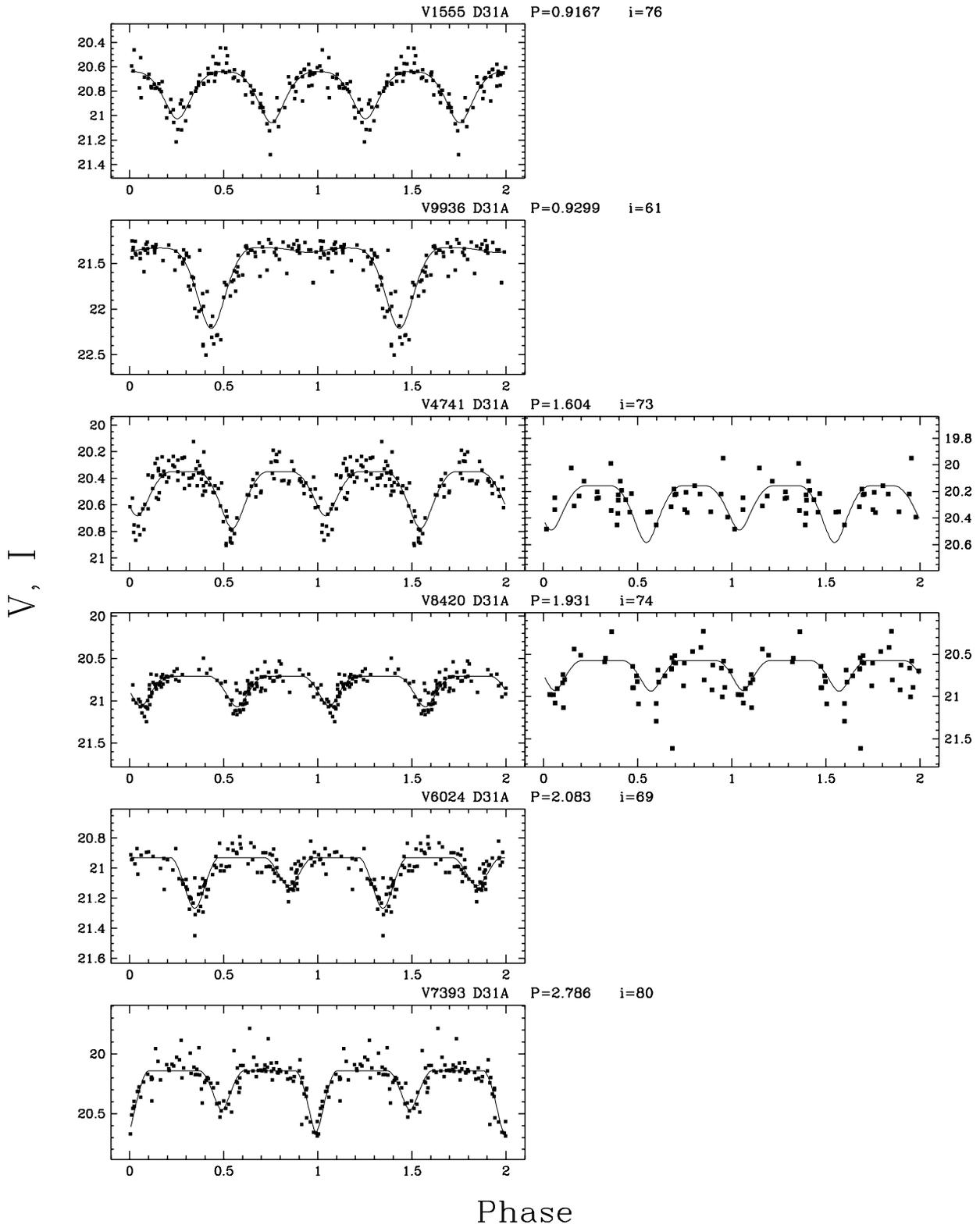}{19.5cm}{0}{83}{83}{-260}{-40}
\caption{$V,I$ lightcurves of eclipsing binaries found in the 
field M31A. The thin continuous line represents for each system
the best fit curve (fitted to the $V$ data).}
\label{fig:ecl1}
\end{figure}
\begin{figure}[p]
\plotfiddle{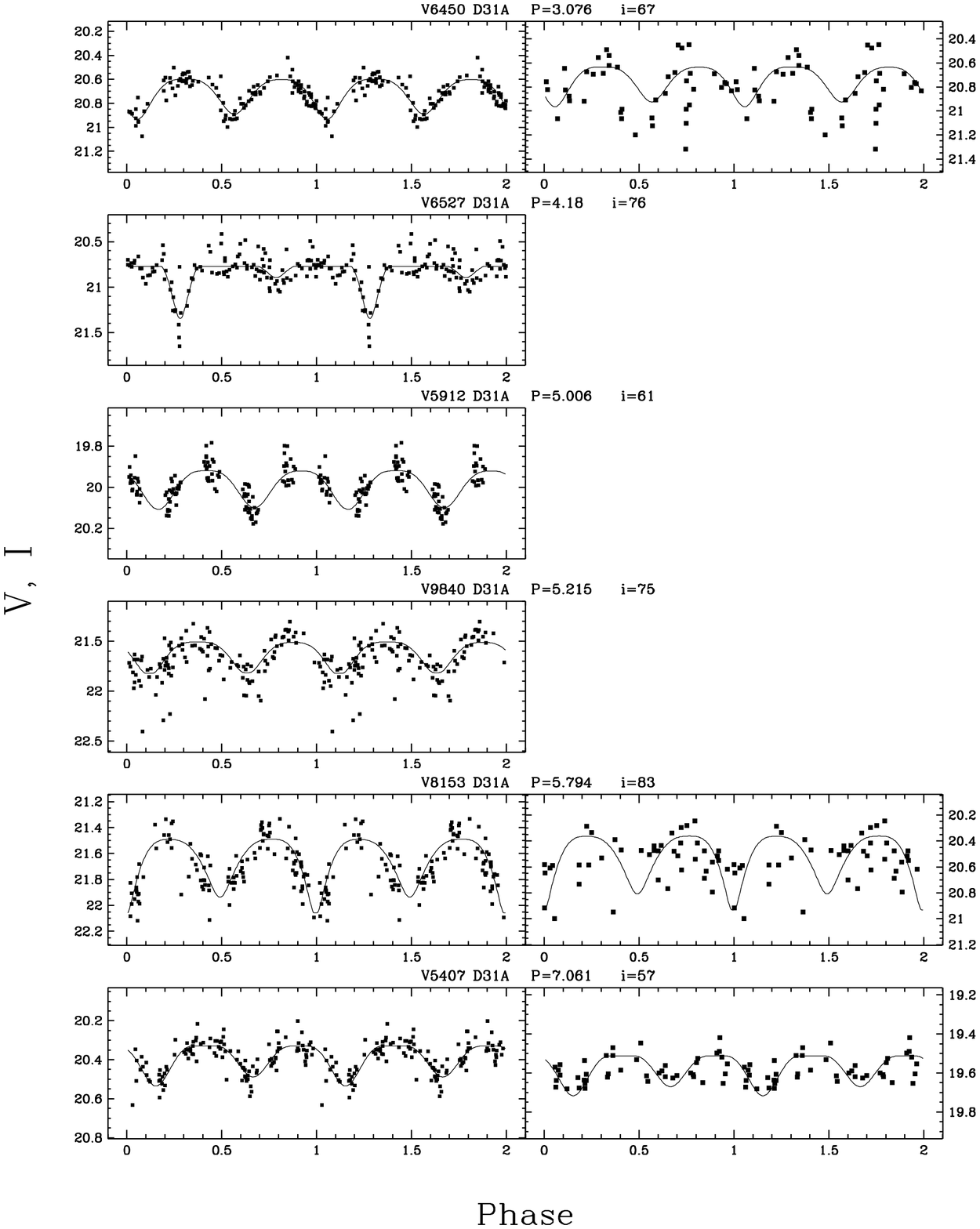}{19.5cm}{0}{83}{83}{-260}{-40}
\caption{Continued from  Fig.\ref{fig:ecl1}.}
\label{fig:ecl2}
\end{figure}
\begin{figure}[t]
\plotfiddle{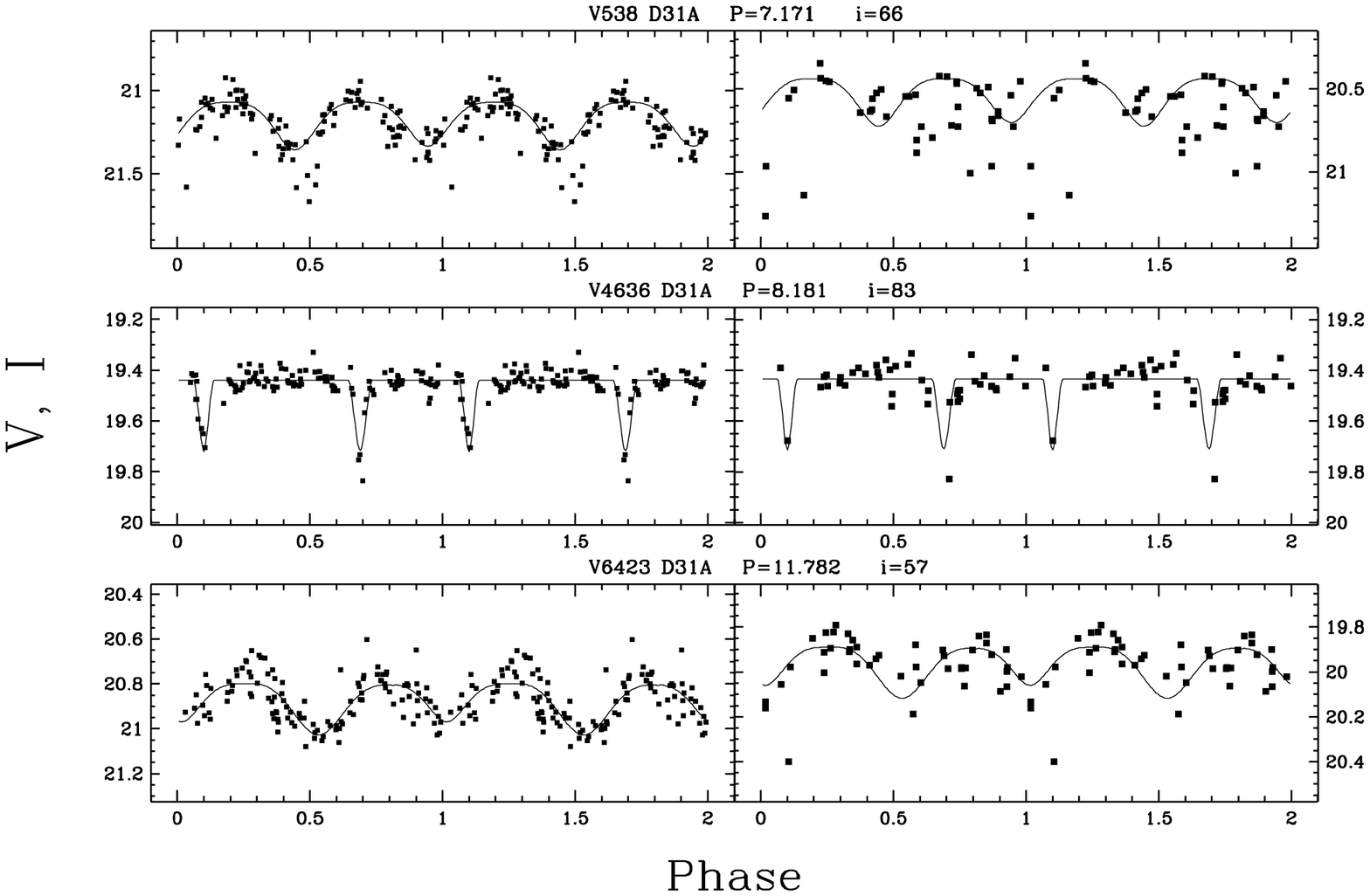}{9.5cm}{0}{83}{83}{-260}{-320}
\caption{Continued from  Fig.\ref{fig:ecl1}.}
\label{fig:ecl3}
\end{figure}

\subsection{Cepheids}

In Table~\ref{table:ceph} we present the parameters of 43 Cepheids in
the M31A field, sorted by the period $P$.  For each Cepheid we present
its name, 2000.0 coordinates, value of the variability index $J_S$,
period $P$, flux-weighted mean magnitudes $\langle V\rangle$ and (when
available) $\langle I\rangle$, and the $V$-band amplitude of the
variation $A$.  In Figs.\ref{fig:ceph1}--\ref{fig:ceph8} we show the
phased $V,I$ lightcurves of our Cepheids. Also shown is the best fit
template lightcurve (Stetson 1996), which was fitted to the $V$ data
and then for the $I$ data only the zero-point offset was allowed.

\subsection{Other periodic variables}

For some of the variables preliminary classified as Cepheids
(Section~5.3.1), we decided upon closer examination to classify them
as ``other periodic variables''.  In Table~\ref{table:per} we present
the parameters of 6 possible periodic variables other than Cepheids
and eclipsing binaries in the M31A field, sorted by the increasing
period $P$.  For each variable we present its name, 2000.0
coordinates, value of the variability index $J_S$, period $P$,
error-weighted mean magnitudes $\bar{V}$ and (when available)
$\bar{I}$. To quantify the amplitude of the variability, we also give
the standard deviations of the measurements in the $V$ and $I$ bands,
$\sigma_{V}$ and $\sigma_{I}$.

\begin{figure}[p]
\plotfiddle{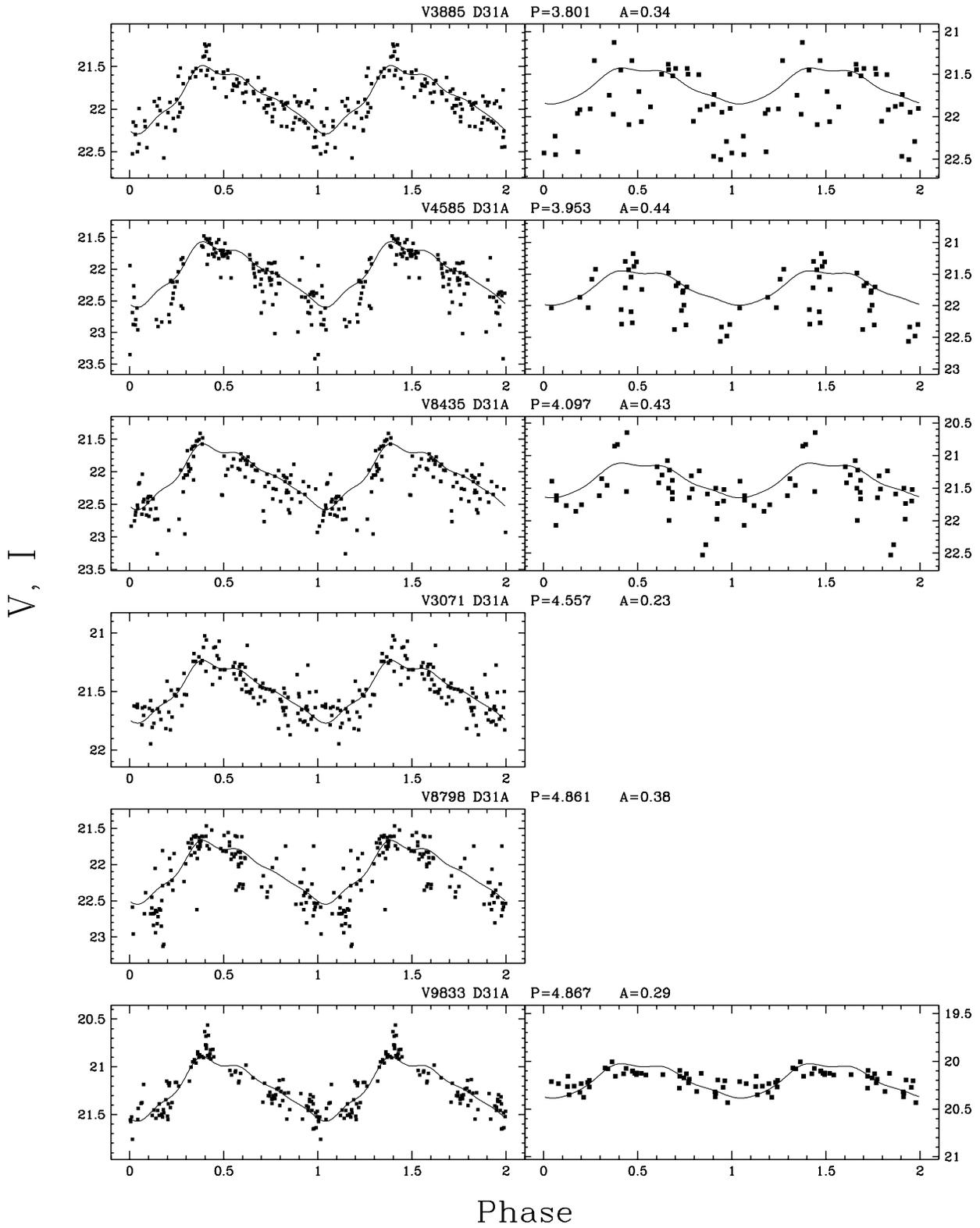}{19.5cm}{0}{83}{83}{-260}{-40}
\caption{$V,I$ lightcurves of Cepheid variables found in the 
field M31A. The thin continuous line represents for each star the best
fit Cepheid template (fitted to the $V$ data). }
\label{fig:ceph1}
\end{figure}
\begin{figure}[p]
\plotfiddle{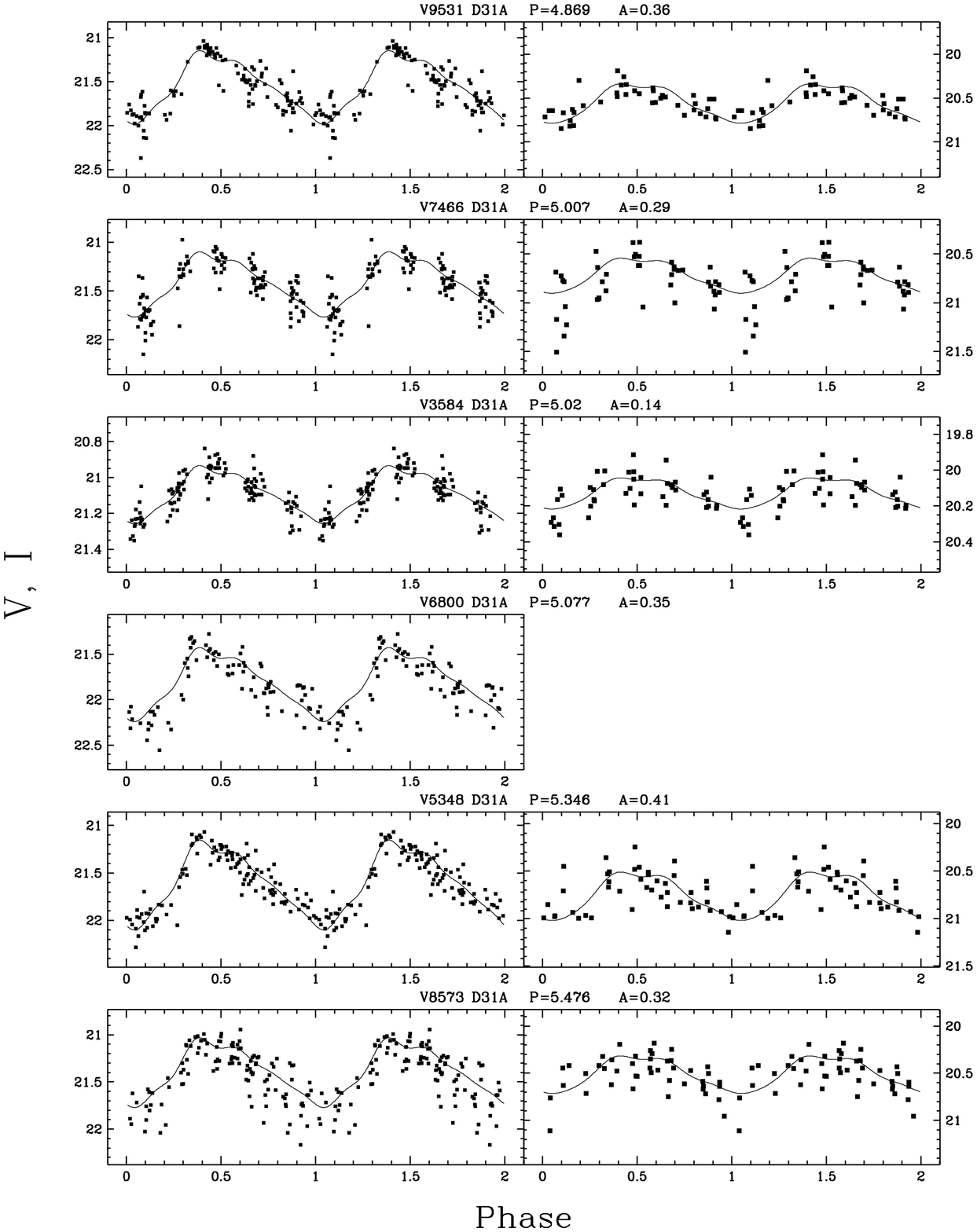}{19.5cm}{0}{83}{83}{-260}{-40}
\caption{Continued from  Fig.\ref{fig:ceph1}.}
\label{fig:ceph2}
\end{figure}
\begin{figure}[p]
\plotfiddle{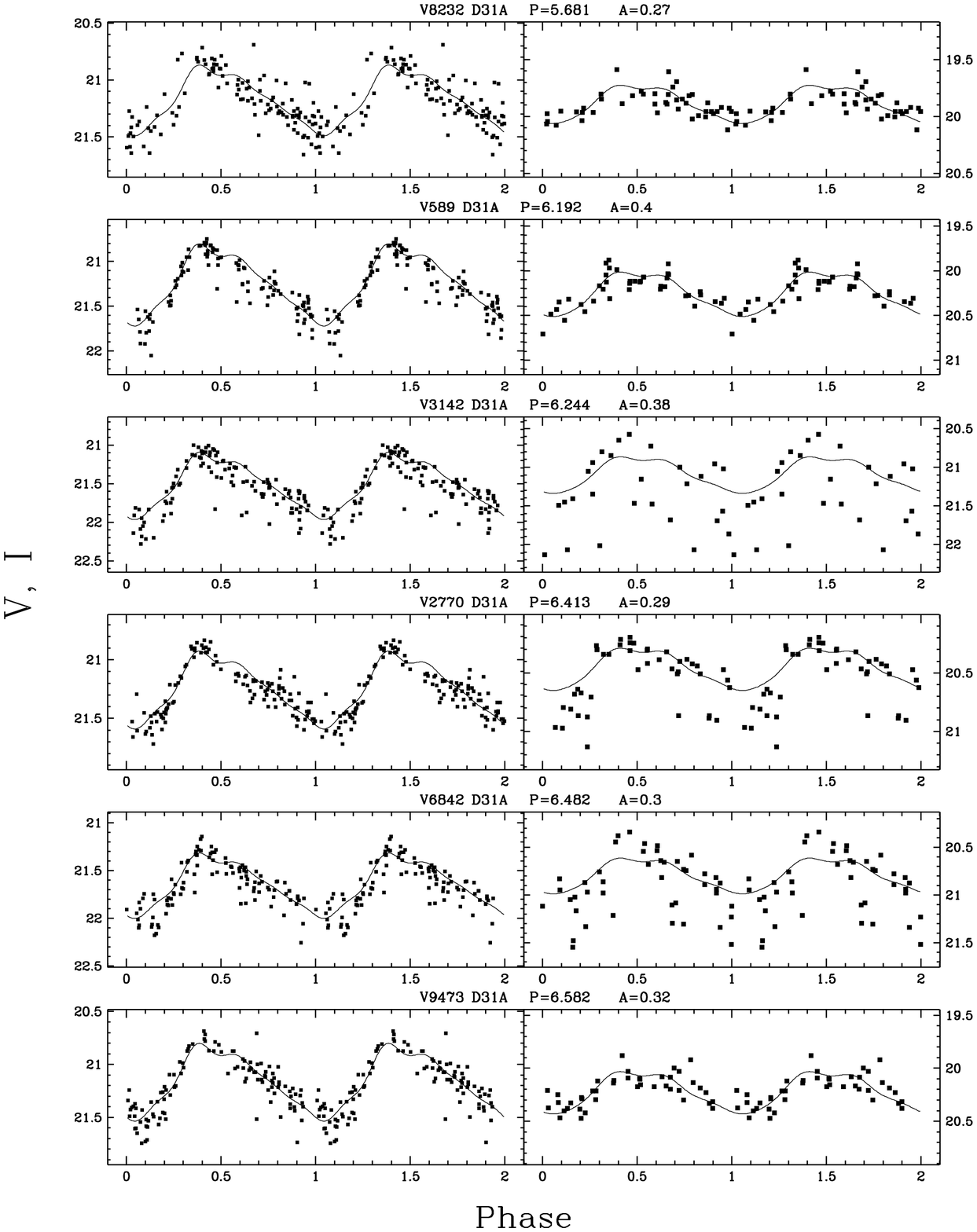}{19.5cm}{0}{83}{83}{-260}{-40}
\caption{Continued from  Fig.\ref{fig:ceph1}.}
\label{fig:ceph3}
\end{figure}
\begin{figure}[p]
\plotfiddle{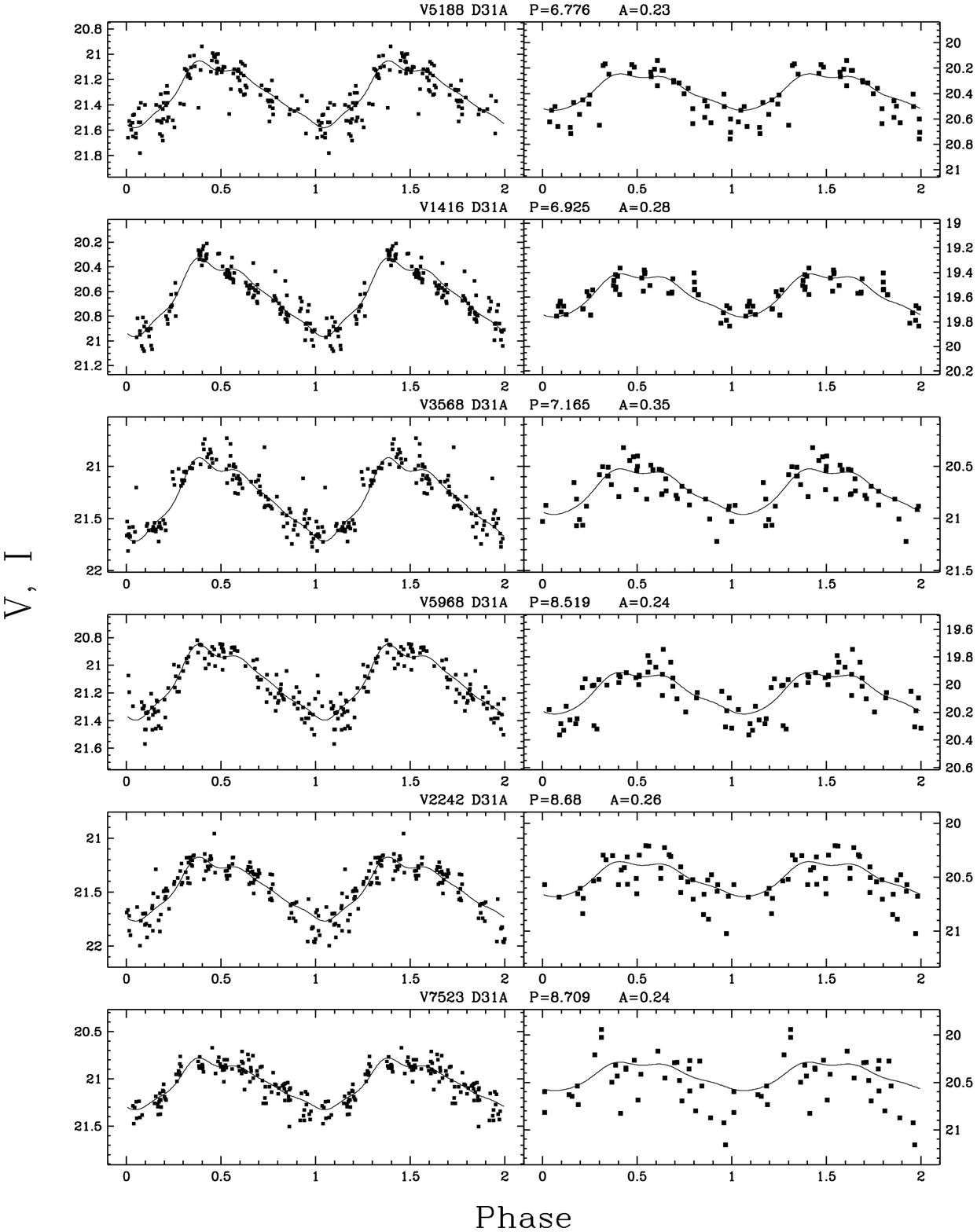}{19.5cm}{0}{83}{83}{-260}{-40}
\caption{Continued from  Fig.\ref{fig:ceph1}.}
\label{fig:ceph4}
\end{figure}
\begin{figure}[p]
\plotfiddle{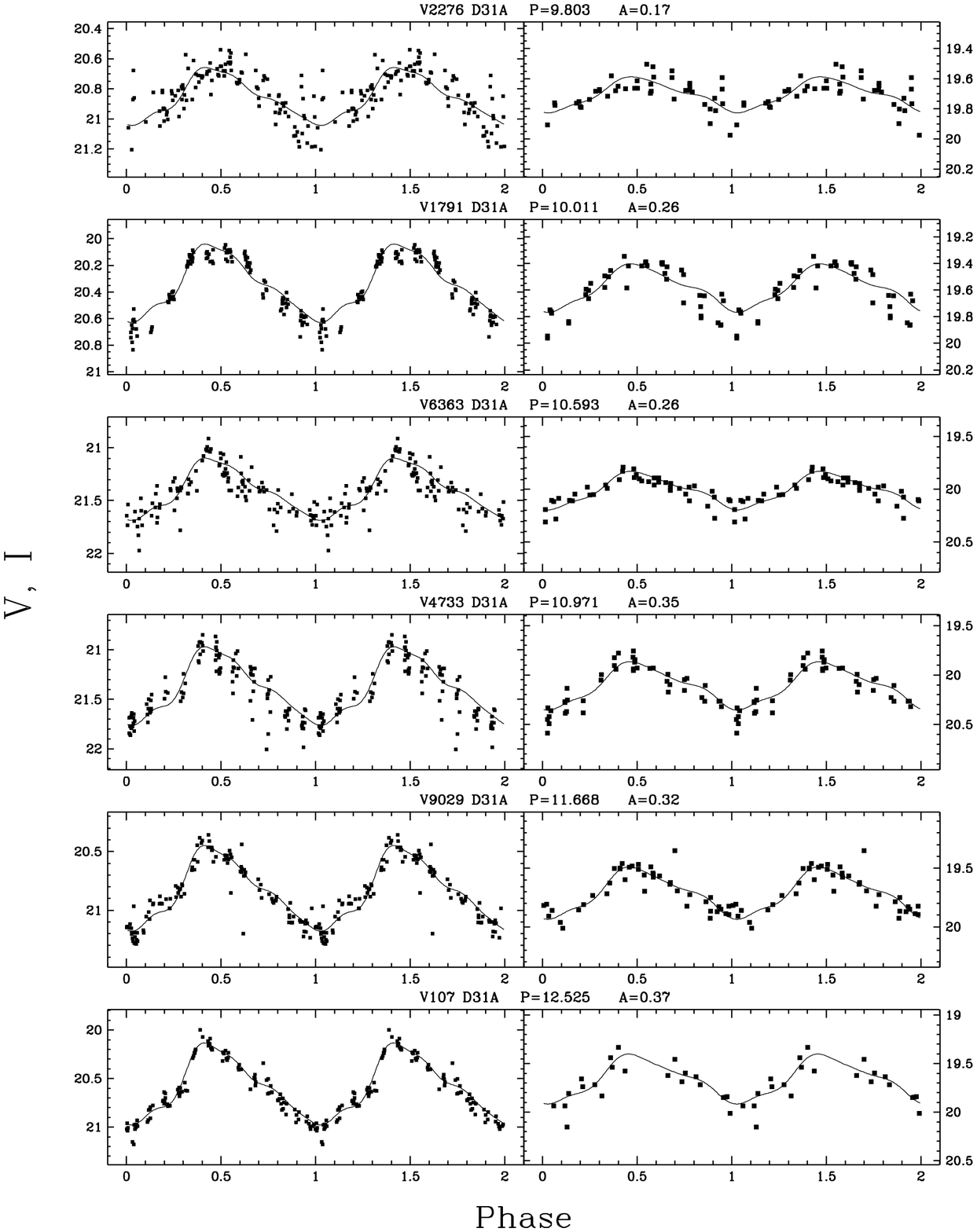}{19.5cm}{0}{83}{83}{-260}{-40}
\caption{Continued from  Fig.\ref{fig:ceph1}.}
\label{fig:ceph5}
\end{figure}
\begin{figure}[p]
\plotfiddle{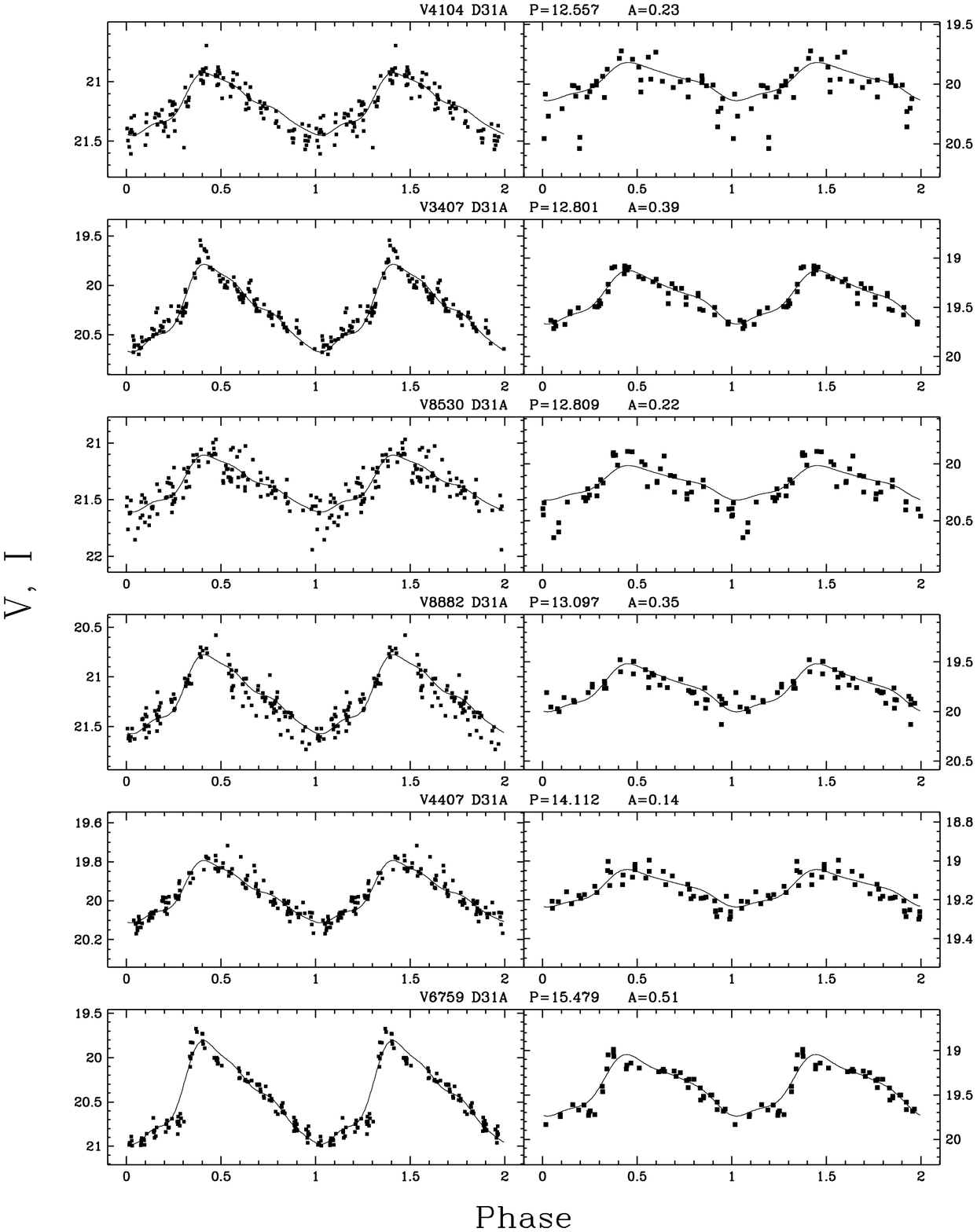}{19.5cm}{0}{83}{83}{-260}{-40}
\caption{Continued from  Fig.\ref{fig:ceph1}.}
\label{fig:ceph6}
\end{figure}
\begin{figure}[p]
\plotfiddle{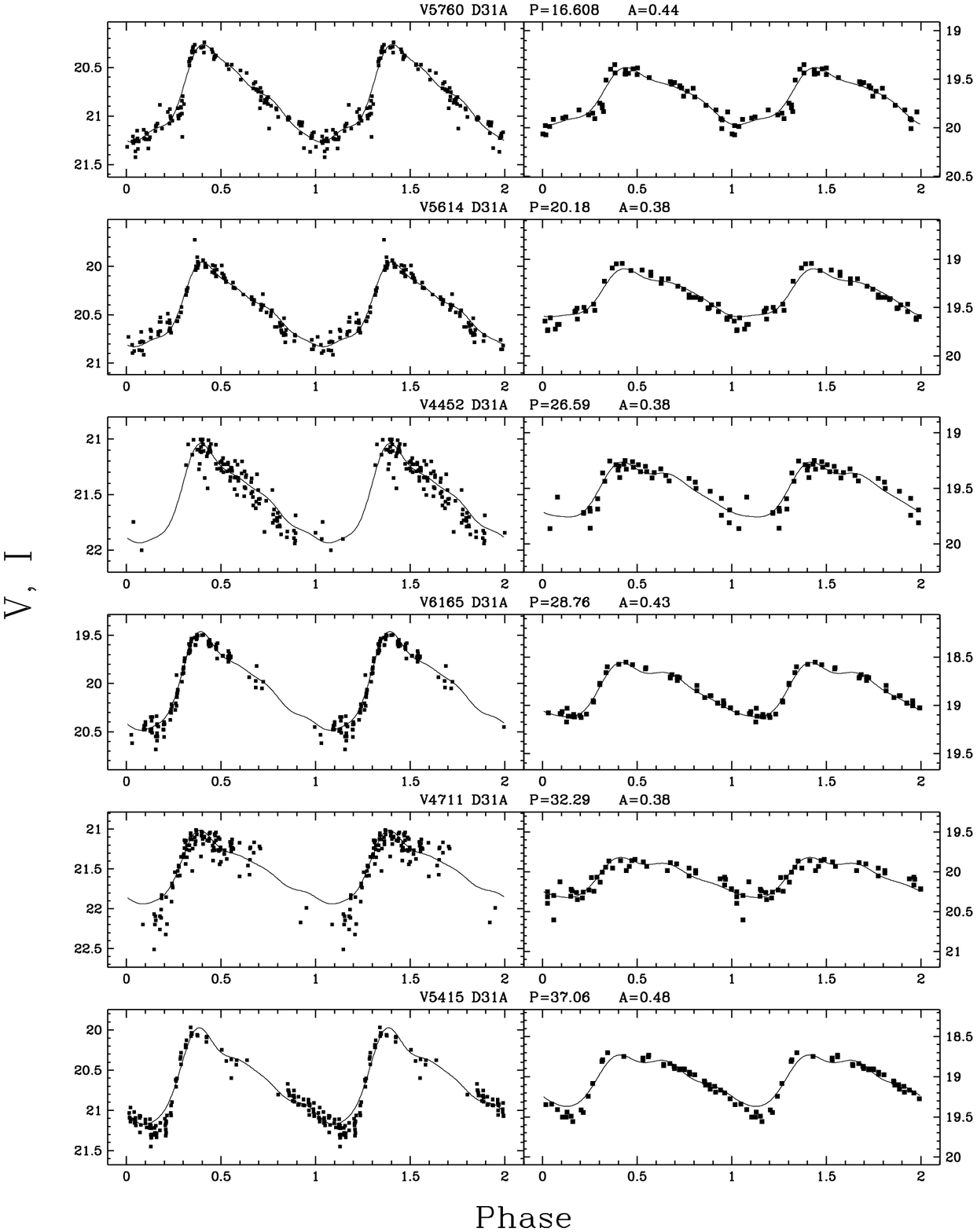}{19.5cm}{0}{83}{83}{-260}{-40}
\caption{Continued from  Fig.\ref{fig:ceph1}.}
\label{fig:ceph7}
\end{figure}
\begin{figure}[p]
\plotfiddle{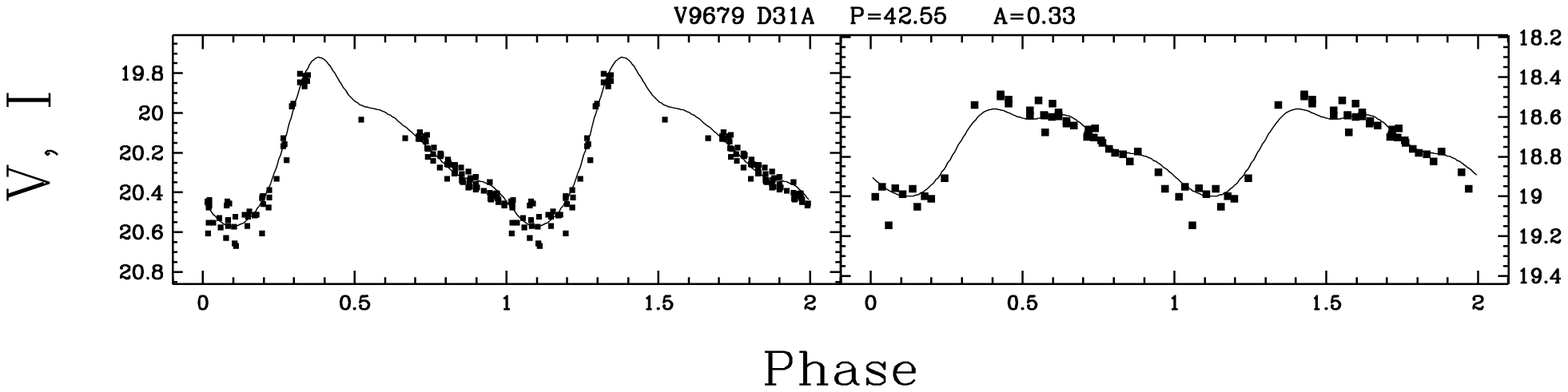}{4cm}{0}{83}{83}{-260}{-520}
\caption{Continued from  Fig.\ref{fig:ceph1}.}
\label{fig:ceph8}
\end{figure}

\tablenum{1} 
\begin{planotable}{lrrrrrrrrcrc}
\tablewidth{42pc}
\tablecaption{DIRECT Eclipsing Binaries in M31A}
\tablehead{ \colhead{Name} & \colhead{$\alpha_{2000.0}$} &
\colhead{$\delta_{2000.0}$} & \colhead{$J_S$} & \colhead{$P$}   &
\colhead{$V_{max}$} & \colhead{$I_{max}$} &
\colhead{$R_1$} & \colhead{$R_2$} & \colhead{$i$} & \colhead{$e$}
& \colhead{Comments} \\ \colhead{(D31A)} & \colhead{$[\deg]$} &
\colhead{$[\deg]$} & \colhead{} & \colhead{$[days]$} & \colhead{} & \colhead{} &
\colhead{} & \colhead{} & \colhead{[deg]}  & \colhead{} & \colhead{} } 
\startdata
V1555 & 11.2717 & 41.6462 & 1.07 & 0.917 & 20.64 &\nodata& 0.59 & 0.41 & 76 & 0.00 & V6913 D31B \nl
V9936 & 11.4523 & 41.7306 & 0.77 & 0.930 & 21.33 &\nodata& 0.65 & 0.34 & 61 & 0.06 & \nl
V4741 & 11.3337 & 41.7511 & 1.43 & 1.604 & 20.35 & 20.16 & 0.51 & 0.40 & 73 & 0.03 & \nl
V8420 & 11.4072 & 41.7188 & 1.21 & 1.931 & 20.71 & 20.58 & 0.42 & 0.37 & 74 & 0.04 & \nl
V6024 & 11.3611 & 41.6841 & 0.83 & 2.083 & 20.93 &\nodata& 0.38 & 0.37 & 69 & 0.00 & \nl
V7393 & 11.3851 & 41.7951 & 0.85 & 2.786 & 20.14 &\nodata& 0.35 & 0.31 & 80 & 0.03 & \nl
V6450 & 11.3681 & 41.7400 & 1.39 & 3.076 & 20.60 & 20.63 & 0.56 & 0.44 & 67 & 0.04 & \nl
V6527 & 11.3671 & 41.8256 & 0.84 & 4.180 & 20.77 &\nodata& 0.31 & 0.31 & 76 & 0.01 & \nl
V5912 & 11.3563 & 41.7510 & 0.88 & 5.006 & 19.92 &\nodata& 0.67 & 0.33 & 61 & 0.01 & \nl
V9840 & 11.4495 & 41.7034 & 0.78 & 5.215 & 21.51 &\nodata& 0.63 & 0.36 & 75 & 0.03 & \nl
V8153 & 11.4026 & 41.6701 & 0.87 & 5.794 & 21.49 & 20.36 & 0.56 & 0.44 & 83 & 0.14 & \nl
V5407 & 11.3464 & 41.7504 & 1.02 & 7.061 & 20.33 & 19.51 & 0.51 & 0.44 & 57 & 0.02 & \nl
 V538 & 11.2420 & 41.6695 & 0.85 & 7.171 & 21.07 & 20.44 & 0.59 & 0.40 & 66 & 0.00 & \nl
V4636 & 11.3321 & 41.7514 & 0.80 & 8.181 & 19.44 & 19.43 & 0.14 & 0.14 & 83 & 0.18 & DEB \nl
V6423 & 11.3671 & 41.7533 & 0.85 &11.782 & 20.80 & 19.89 & 0.54 & 0.46 & 57 & 0.15 &
\enddata 
\tablecomments{Variable V4636 D31A with period $P=8.181\;days$ is a detached 
eclipsing binary (DEB) with significant ellipticity. V1555 D31A was found
as V6913 D31B in Paper I, with identical period, $V_{max}=20.63$
and  $I_{max}=20.06$.}
\label{table:ecl}
\end{planotable}

\tablenum{2} 
\begin{planotable}{lrrrrrrrc}
\tablewidth{35pc}
\tablecaption{DIRECT Cepheids in M31A}
\tablehead{ \colhead{Name} & \colhead{$\alpha_{2000.0}$} &
\colhead{$\delta_{2000.0}$} & \colhead{$J_S$} & \colhead{$P$}  &
\colhead{$\langle V\rangle$} & \colhead{$\langle I\rangle$} & \colhead{$A$} 
& \colhead{Comments} \\ \colhead{(D31A)} & \colhead{[deg]} &
\colhead{[deg]} & \colhead{} & \colhead{$[days]$} & \colhead{} &
\colhead{} & \colhead{} & \colhead{} } 
\startdata 
V3885 & 11.3176 & 41.7868 & 1.13 &  3.801  & 21.87 & 21.63 & 0.34 & \nl
V4585 & 11.3337 & 41.6685 & 1.18 &  3.953  & 22.06 & 21.71 & 0.44 & \nl
V8435 & 11.4083 & 41.6933 & 1.05 &  4.097  & 22.05 & 21.37 & 0.43 & \nl
V3071 & 11.3036 & 41.6866 & 1.05 &  4.557  & 21.49 &\nodata& 0.23 & \nl
V8798 & 11.4135 & 41.8081 & 1.36 &  4.861  & 22.09 &\nodata& 0.38 & \nl
V9833 & 11.4461 & 41.7914 & 1.60 &  4.867  & 21.22 & 20.20 & 0.29 & \nl
V9531 & 11.4361 & 41.7111 & 1.69 &  4.869  & 21.55 & 20.55 & 0.36 & \nl
V7466 & 11.3867 & 41.7830 & 1.15 &  5.007  & 21.42 & 20.71 & 0.29 & \nl
V3584 & 11.3125 & 41.7674 & 0.80 &  5.020  & 21.09 & 20.12 & 0.14 & \nl
V6800 & 11.3780 & 41.6476 & 0.91 &  5.077  & 21.82 &\nodata& 0.35 & \nl
V5348 & 11.3479 & 41.6677 & 1.65 &  5.346  & 21.61 & 20.75 & 0.41 & \nl
V8573 & 11.4116 & 41.7128 & 1.35 &  5.476  & 21.39 & 20.50 & 0.32 & Ma97 125 \nl
V8232 & 11.4011 & 41.7727 & 1.37 &  5.681  & 21.17 & 19.88 & 0.27 & Ma97 124 \nl
 V589 & 11.2430 & 41.6840 & 2.42 &  6.192  & 21.25 & 20.25 & 0.40 & \nl
V3142 & 11.3047 & 41.7006 & 1.83 &  6.244  & 21.52 & 21.09 & 0.38 & \nl
V2770 & 11.2952 & 41.7592 & 1.62 &  6.413  & 21.25 & 20.46 & 0.29 & \nl
V6842 & 11.3770 & 41.7058 & 1.20 &  6.482  & 21.65 & 20.79 & 0.30 & \nl
V9473 & 11.4337 & 41.7366 & 1.76 &  6.582  & 21.16 & 20.22 & 0.32 & Ma97 127 \nl
V5188 & 11.3438 & 41.7022 & 1.52 &  6.776  & 21.31 & 20.38 & 0.23 & Ma97 111 \nl
V1416 & 11.2661 & 41.6983 & 2.98 &  6.925  & 20.64 & 19.57 & 0.28 & \nl
V3568 & 11.3117 & 41.7760 & 2.10 &  7.165  & 21.31 & 20.73 & 0.35 & \nl
V5968 & 11.3586 & 41.7275 & 1.64 &  8.519  & 21.12 & 20.05 & 0.24 & Ma97 114 \nl
V2242 & 11.2854 & 41.7508 & 1.12 &  8.680  & 21.47 & 20.51 & 0.26 & \nl
V7523 & 11.3917 & 41.6581 & 1.82 &  8.709  & 21.05 & 20.43 & 0.24 & Ma97 121 \nl
V2276 & 11.2886 & 41.6656 & 1.42 &  9.803  & 20.86 & 19.70 & 0.17 & V7553 D31B \nl
V1791 & 11.2763 & 41.6948 & 3.04 & 10.011  & 20.35 & 19.58 & 0.26 & \nl
V6363 & 11.3683 & 41.6595 & 1.18 & 10.593  & 21.41 & 20.01 & 0.26 & Ma97 117 \nl
V4733 & 11.3324 & 41.7888 & 1.99 & 10.971  & 21.39 & 20.11 & 0.35 & \nl
V9029 & 11.4211 & 41.7472 & 2.95 & 11.668  & 20.84 & 19.71 & 0.32 & Ma97 126 \nl
 V107 & 11.2255 & 41.7051 & 4.41 & 12.525  & 20.59 & 19.66 & 0.37 & \nl
V4104 & 11.3261 & 41.6504 & 1.45 & 12.557  & 21.21 & 19.98 & 0.23 & \nl
V3407 & 11.3085 & 41.7645 & 4.39 & 12.801  & 20.27 & 19.40 & 0.39 & \nl
V8530 & 11.4076 & 41.8088 & 0.79 & 12.809  & 21.38 & 20.17 & 0.22 & \nl
V8882 & 11.4152 & 41.8081 & 1.51 & 13.097  & 21.21 & 19.76 & 0.35 & \nl
V4407 & 11.3308 & 41.6660 & 1.70 & 14.112  & 19.97 & 19.14 & 0.14 & \nl
V6759 & 11.3772 & 41.6576 & 6.45 & 15.479  & 20.46 & 19.41 & 0.51 & Ma97 118 \nl
V5760 & 11.3544 & 41.7088 & 4.49 & 16.608  & 20.83 & 19.69 & 0.44 & \nl
V5614 & 11.3509 & 41.7351 & 5.43 & 20.18~~ & 20.46 & 19.37 & 0.38 & Ma97 113 \nl
V4452 & 11.3319 & 41.6506 & 1.37 & 26.59~~ & 21.56 & 19.52 & 0.38 & \nl
V6165 & 11.3632 & 41.6997 & 5.58 & 28.76~~ & 20.05 & 18.84 & 0.43 & Ma97 116 \nl
V4711 & 11.3361 & 41.6596 & 1.15 & 32.29~~ & 21.54 & 20.06 & 0.38 & \nl
V5415 & 11.3483 & 41.6933 & 4.46 & 37.06~~ & 20.64 & 19.02 & 0.48 & Ma97 112 \nl
V9679 & 11.4405 & 41.7749 & 2.46 & 42.55~~ & 20.18 & 18.75 & 0.33 & Ma97 129
\enddata
\label{table:ceph}
\tablecomments{Variable V2276 D31A (Ma97 108) was found as V7553 D31B
in Paper I, with period $P=9.482\;days$, $\langle V\rangle = 20.93$
and $\langle I\rangle = 19.77$. }
\end{planotable}

\subsection{Miscellaneous  variables}	
	
In Table~\ref{table:misc} we present the parameters of 11 miscellaneous
variables in the M31A field, sorted by the decreasing value of the
mean magnitude $\bar{V}$. For each variable we present its name,
2000.0 coordinates, value of the variability index $J_S(>1.2)$, mean
magnitudes $\bar{V}$ and $\bar{I}$.  To quantify the amplitude of the
variability, we also give the standard deviations of the measurements
in $V$ and $I$ bands, $\sigma_{V}$ and $\sigma_{I}$.  In the
``Comments'' column we give a rather broad sub-classification of the
variability: LP -- possible long-period variable ($P>56\;days$); IRR
-- irregular variable. Unlike the M31B field (Paper I), all of the
miscellaneous variables seem to represent the LP type of variability.

\tablenum{3} 
\begin{planotable}{lllllllllc}
\tablewidth{38pc}
\tablecaption{DIRECT Other Periodic Variables in M31A}
\tablehead{ \colhead{Name} & \colhead{$\alpha_{2000.0}$} &
\colhead{$\delta_{2000.0}$} & \colhead{$J_S$} & \colhead{$P$}   &
\colhead{$\bar{V}$} & \colhead{$\bar{I}$} &
\colhead{$\sigma_V$} & \colhead{$\sigma_I$} & \colhead{Comments} \\
\colhead{(D31A)} &  \colhead{[deg]} &  \colhead{[deg]} &
\colhead{} & \colhead{$[days]$} & \colhead{} &
\colhead{} & \colhead{} & \colhead{} & \colhead{} }
\startdata
V1494 & 11.2659 & 41.7755 & 1.30 & 23.15 & 21.56 &\nodata& 0.33 &\nodata& \nl
 V410 & 11.2355 & 41.7199 & 1.10 & 28.62 & 20.92 &\nodata& 0.13 &\nodata& \nl
V1911 & 11.2800 & 41.6786 & 1.68 & 36.74 & 19.99 & 19.25 & 0.12 & 0.14  & RV Tau \nl
V3368 & 11.3104 & 41.6779 & 2.62 & 46.18 & 21.14 & 20.34 & 0.36 & 0.23  & RV Tau \nl
V2977 & 11.3018 & 41.6835 & 1.27 & 53.3  & 21.21 & 20.13 & 0.18 & 0.28  & RV Tau \nl
V9659 & 11.4421 & 41.6874 & 1.75 & 56.6  & 20.77 & 19.96 & 0.40 & 0.27  & RV Tau 
\enddata
\label{table:per}
\end{planotable}

\tablenum{4} 
\begin{planotable}{llllllllc}
\tablewidth{35pc}
\tablecaption{DIRECT Miscellaneous Variables in M31A}
\tablehead{ \colhead{Name} & \colhead{$\alpha_{2000.0}$} &
\colhead{$\delta_{2000.0}$} & \colhead{$J_S$} &
\colhead{$\bar{V}$} & \colhead{$\bar{I}$} &
\colhead{$\sigma_V$} & \colhead{$\sigma_I$} & \colhead{Comments} \\
\colhead{(D31A)} & \colhead{[deg]} & \colhead{[deg]} &
\colhead{} & \colhead{} & \colhead{} &
\colhead{} & \colhead{} & \colhead{} }
\startdata
V3222 & 11.3081 & 41.6494 & 1.40 & 19.11 & 17.21 & 0.09 & 0.04 & V8123 D31B \nl 
V3901 & 11.3216 & 41.6633 & 1.61 & 19.25 & 16.72 & 0.10 & 0.05 & LP \nl 
V2109 & 11.2844 & 41.6975 & 1.67 & 20.07 & 19.13 & 0.11 & 0.07 & LP \nl 
V7718 & 11.3926 & 41.7313 & 2.02 & 20.23 & 19.59 & 0.15 & 0.10 & LP \nl 
V8415 & 11.4050 & 41.7867 & 1.98 & 20.37 & 19.65 & 0.17 & 0.12 & LP \nl 
V5764 & 11.3563 & 41.6507 & 1.36 & 20.75 & 20.20 & 0.14 & 0.12 & LP \nl 
V2570 & 11.2912 & 41.7629 & 1.82 & 21.42 & 19.60 & 0.53 & 0.20 & LP \nl 
 V541 & 11.2424 & 41.6594 & 2.47 & 21.51 & 19.20 & 0.54 & 0.13 & V5897 D31B \nl 
  V11 & 11.2240 & 41.6504 & 1.27 & 21.72 & 19.31 & 0.47 & 0.12 & V5075 D31B \nl 
V9459 & 11.4322 & 41.7689 & 1.71 & 21.84 & 19.23 & 0.54 & 0.18 & LP \nl  
V1032 & 11.2565 & 41.6655 & 1.72 & 21.88 & 19.43 & 0.43 & 0.14 & LP 
\enddata
\label{table:misc}
\tablecomments{Variables V3222, V541 and V11 were also found in Paper I.}
\end{planotable}

\begin{figure}[p]
\plotfiddle{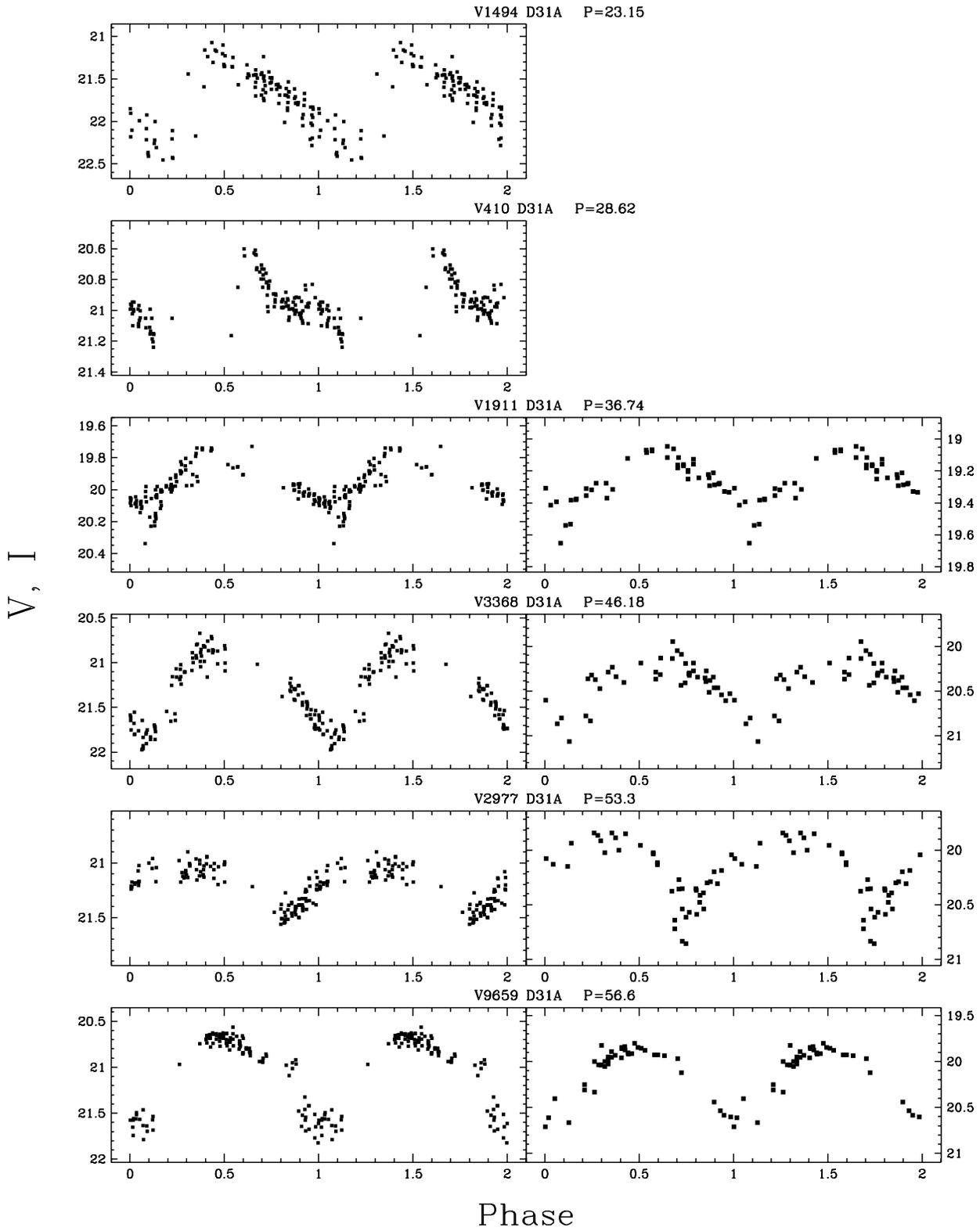}{19.5cm}{0}{83}{83}{-260}{-40}
\caption{$V,I$ lightcurves of other periodic  variables found in the 
field M31A.}
\label{fig:per1}
\end{figure}
 
\begin{figure}[p]
\plotfiddle{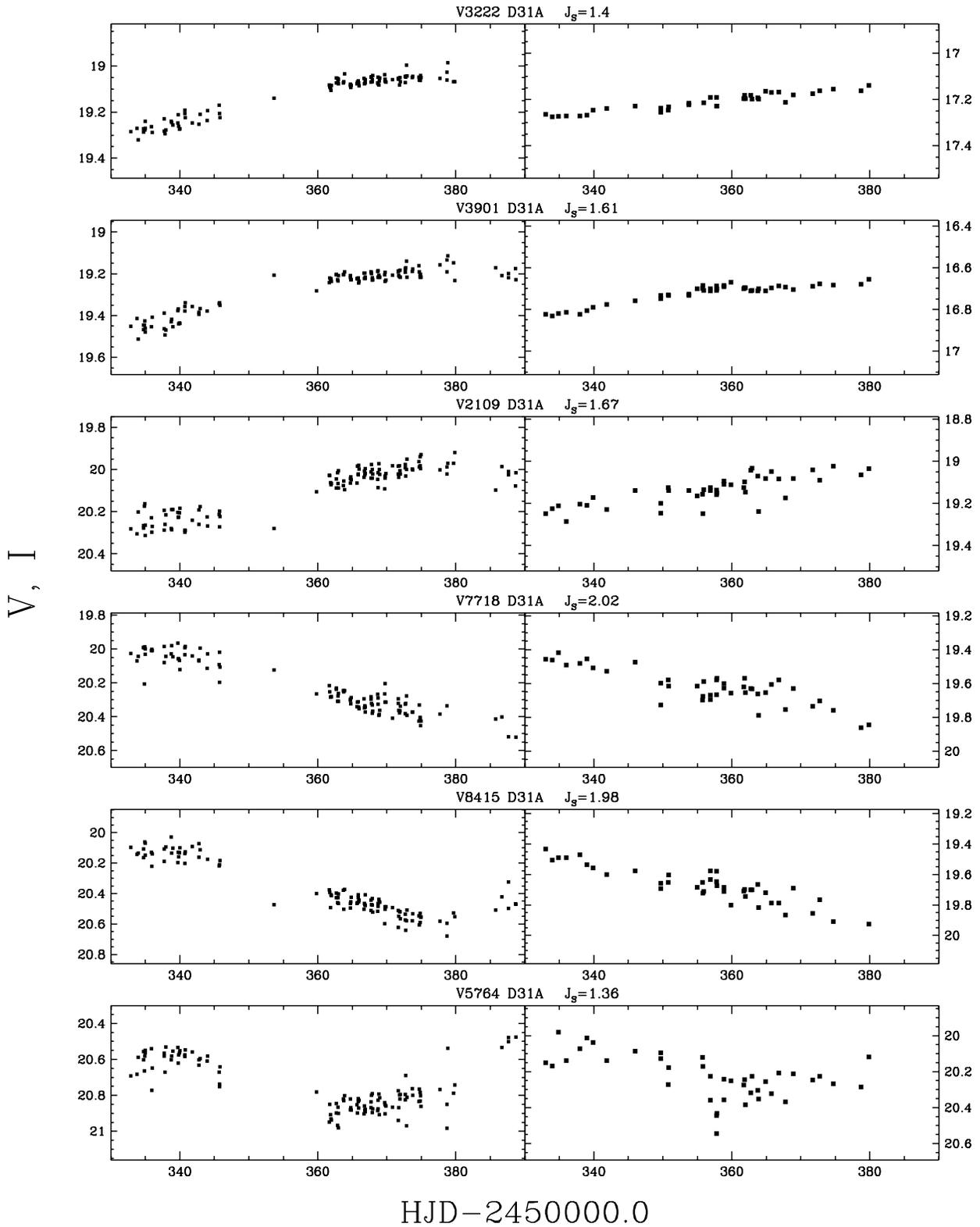}{19.5cm}{0}{83}{83}{-260}{-40}
\caption{$V,I$ lightcurves of miscellaneous variables found in the 
field M31A.}
\label{fig:misc1}
\end{figure}
\begin{figure}[p]
\plotfiddle{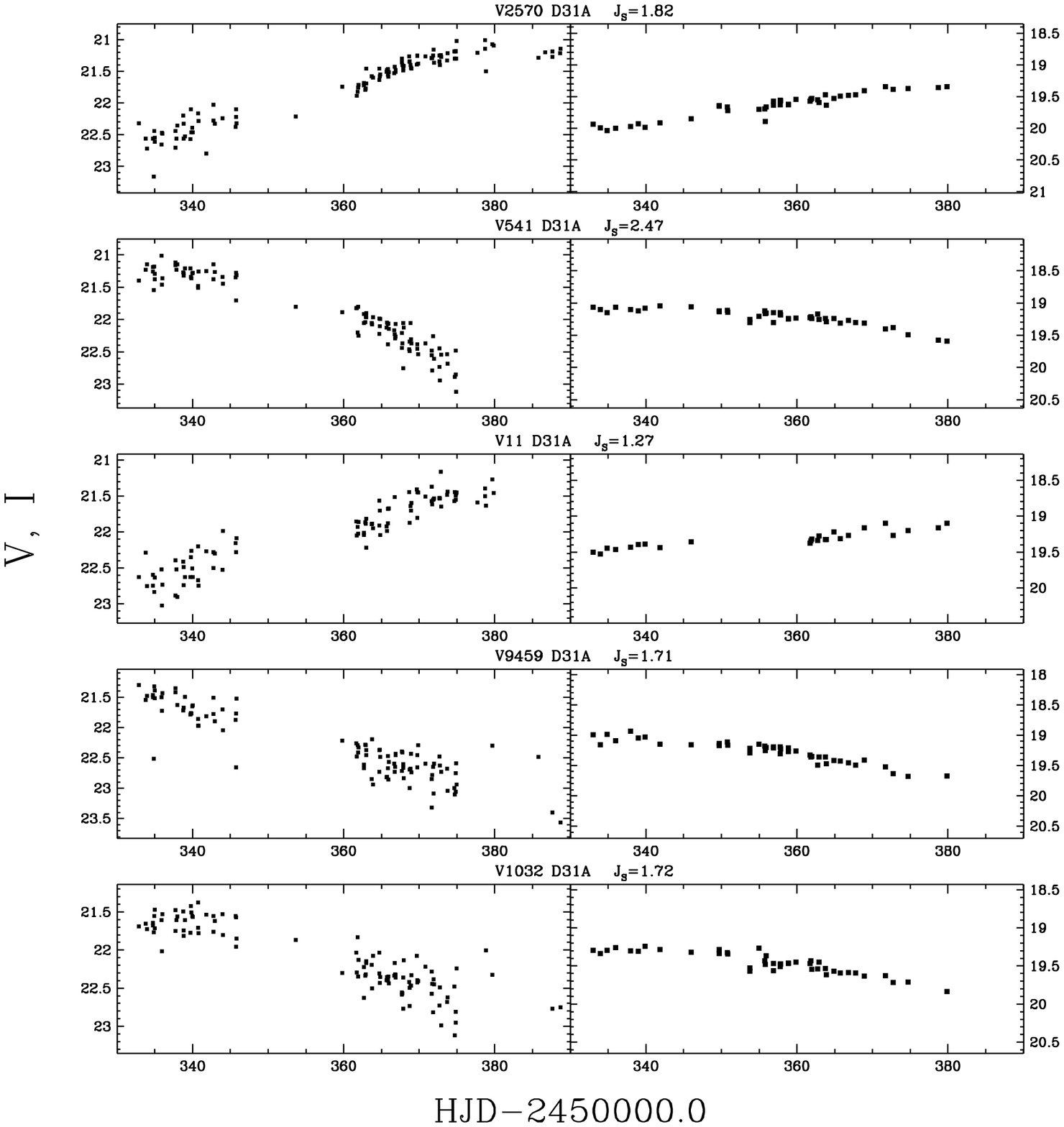}{19.5cm}{0}{83}{83}{-260}{-40}
\caption{Continued from  Fig.\ref{fig:misc1}.}
\label{fig:misc2}
\end{figure}

\subsection{Comparison with other catalogs}

The area of M31A field has not been observed frequently before and the
only overlapping variable star catalog is given by Magnier et
al.~(1997, hereafter Ma97). Of 17 variable stars in Ma97 which are in
our M31A field, we cross-identified 16. Of these 16 stars, one (Ma97
120) we did not classify as variable ($J_S=0.22$) and one (Ma97 123)
was initially classified as variable ($J_S=1.1$), but then failed to
classify as a Cepheid. The remaining 14 stars we classified as
Cepheids (see Table~\ref{table:ceph} for cross-identifications).

There was also by design a slight overlap between the M31A and M31B
fields (Fig.\ref{fig:xy}). Out of two eclipsing binaries found in the
overlap from the M31A field, V1555 D31A was cross-identified as V6913
D31B, with very similar properties of the light curves (see
Table~\ref{table:ecl}).  V538 D31A turned out to have $J_S=0.749$ in
the M31B database, i.e. it very narrowly escaped classification as a
candidate variable star in this field. There was only one Cepheid from
the M31A field in the overlap region, V2276 D31A (also Ma97 108), and
it was cross-identified as V7553 D31B, again with very similar
properties of the light curves (see Table~\ref{table:ceph}). There was
a second Cepheid in the overlap found in Paper I, V7845 D31B
($J_S=0.88$), which failed to qualify as a Cepheid in the M31A field
because of the more stringent requirements for the reduced
$\chi^2/N_{DOF}$ of the fit (Section~5).  We also cross-identified
three miscellaneous variables (see Table~\ref{table:misc}), out of
four detected in the M31A field and five detected in the M31B field,
which fell into the overlap region.

\section{Discussion}

\begin{figure}[t]
\plotfiddle{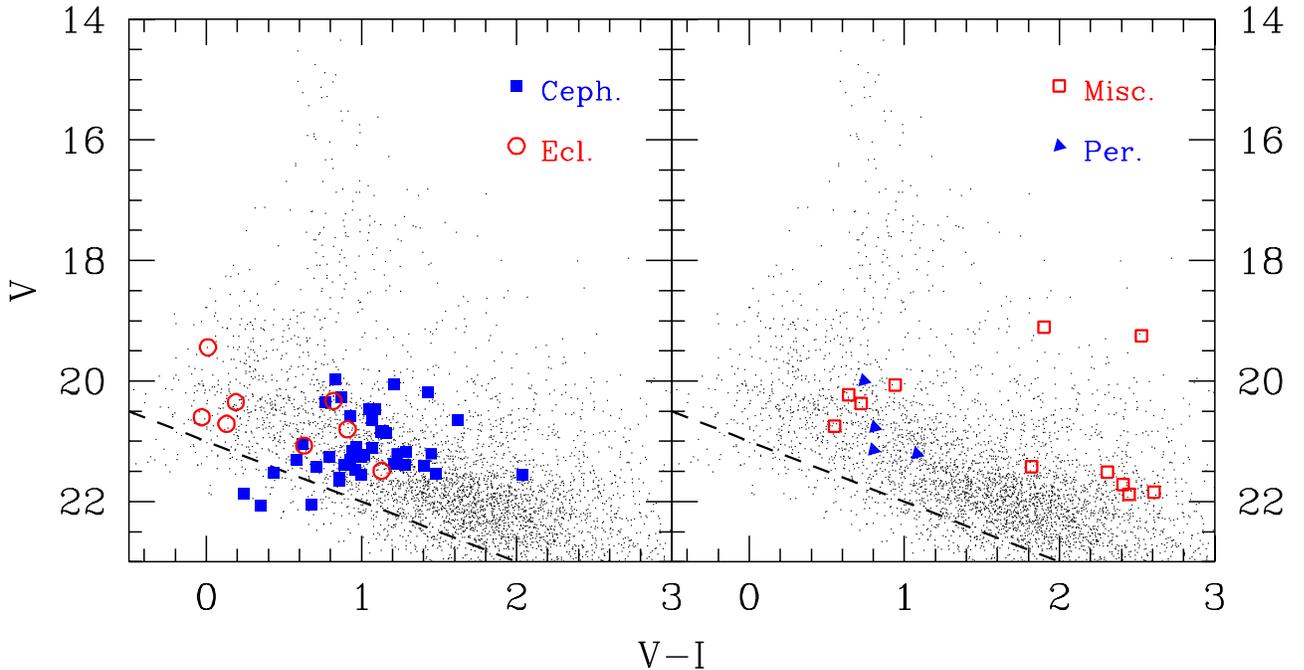}{8cm}{0}{85}{85}{-250}{-340}
\caption{$V,\;V-I$ color-magnitude diagrams for the variable stars
found in the field M31A. The eclipsing binaries and Cepheids are
plotted in the left panel and the other periodic variables and
miscellaneous variables are plotted in the right panel. The dashed
line corresponds to the $I$ detection limit of $I\sim21\;{\rm mag}$.
\label{fig:cmd}}
\end{figure}

In Fig.\ref{fig:cmd} we show $V, V-I$ color-magnitude diagrams for
the variable stars found in the field M31A. The eclipsing binaries and
Cepheids are plotted in the left panel and the other periodic
variables and miscellaneous variables are plotted in the right panel.
As expected, some of the eclipsing binaries occupy the blue upper main
sequence of M31 stars, but there is a group of eclipsing binaries with
$V-I\sim1.0$.  The Cepheid variables group near $V-I\sim1.0$, with the
exception of possibly highly reddened system V4452 D31A. The other
periodic variable stars have positions on the CMD similar to the
Cepheids. The miscellaneous variables are scattered throughout the CMD
and might represent few classes of variability, but most of them are
red with $V-I=1.8-2.6$, and are probably Mira variables.

\begin{figure}[t]
\plotfiddle{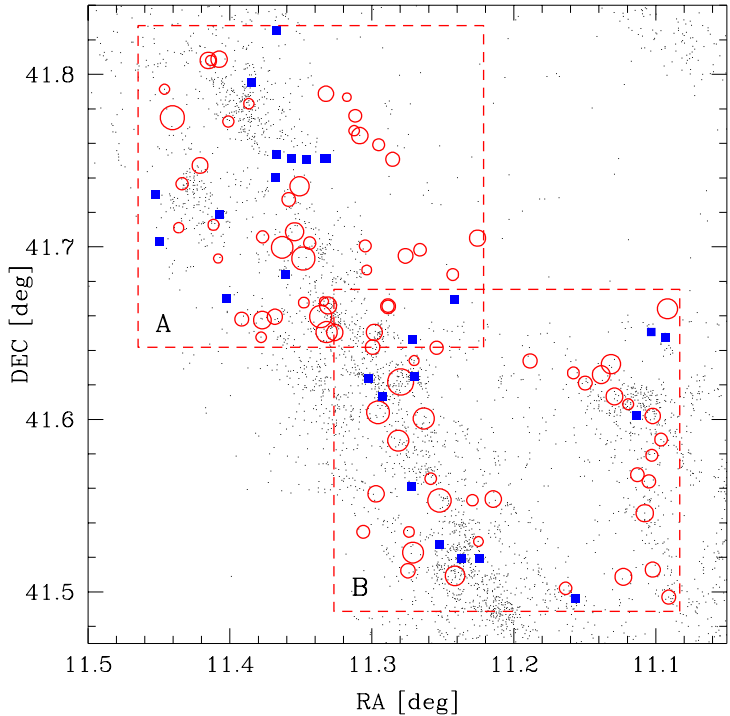}{8cm}{0}{50}{50}{-160}{-85}
\caption{Location of eclipsing binaries (filled squares) and Cepheids
(open circles) in the fields M31A and M31B, along with the blue stars
($B-V<0.4$) selected from the photometric survey of M31 by Magnier et
al.~(1992) and Haiman et al.~(1994). The sizes of the circles
representing the Cepheids variables are proportional to the logarithm
of their period.
\label{fig:xy}}
\end{figure}

In Fig.\ref{fig:xy} we plot the location of eclipsing binaries and
Cepheids in the fields M31A and M31B, along with the blue stars
($B-V<0.4$) selected from the photometric survey of M31 by Magnier et
al.~(1992) and Haiman et al.~(1994). The sizes of the circles
representing the Cepheids variables are proportional to the logarithm
of their period. As could have been expected, both types of variables
group along the spiral arms, as they represent relatively young
populations of stars. However, in the field M31A there is a group of
Cepheids located outside the population of the blue stars (at $\sim
11.3,41.76\;\deg$ and below). Also, these Cepheids as well as the Cepheids
located in the inner spiral arm of M31 (field M31B at RA$\sim 11.1\;\deg$)
have on average shorter periods than the Cepheids located in the outer
spiral arm. We will explore various properties of our sample of
Cepheids in the future paper (Sasselov et al.~1998, in preparation).

\acknowledgments{We would like to thank the TAC of the
Michigan-Dartmouth-MIT (MDM) Observatory and the TAC of the
F.~L.~Whipple Observatory (FLWO) for the generous amounts of telescope
time devoted to this project. We are very grateful to Bohdan
Paczy\'nski for motivating us to undertake this project and his always
helpful comments and suggestions.  Przemek Wo\'zniak supplied us with
FITS-manipulation programs we use to create the finding charts. The
staff of the MDM and the FLWO observatories is thanked for their
support during the long observing runs.  JK was supported by NSF grant
AST-9528096 to Bohdan Paczy\'nski and by the Polish KBN grant
2P03D011.12. KZS was supported by the Harvard-Smithsonian Center for
Astrophysics Fellowship.  JLT was supported by the NSF grant
AST-9401519.}

\end{document}